\documentclass[journal abbreviation]{copernicus}
\usepackage{amsmath}
\usepackage{subfig}
\usepackage{graphicx}
\usepackage[font=small,labelfont=bf]{caption}
\usepackage{blindtext}
\usepackage{cuted}

\begin{document}


\title{Planar magnetic structures in coronal mass ejection-driven sheath regions}

\author[1]{Erika Palmerio}
\author[1]{Emilia K. J. Kilpua}
\author[2,3]{Neel P. Savani}

\affil[1]{Department of Physics, P.O. Box 64, University of Helsinki, Finland}
\affil[2]{Goddard Planetary Heliophysics Institute (GPHI), University of Maryland, Baltimore County, Maryland, USA}
\affil[3]{NASA Goddard Space Flight Center, Greenbelt MD 20771, USA}


\runningtitle{PMSs in CME sheaths}

\runningauthor{Palmerio et al.}

\correspondence{Erika Palmerio\\ (erika.palmerio@helsinki.fi)}

\received{}
\pubdiscuss{} 
\revised{}
\accepted{}
\published{}


\firstpage{1}

\maketitle  

\begin{abstract}
Planar magnetic structures (PMSs) are periods in the solar wind during which interplanetary magnetic field vectors are nearly parallel to a single plane. One of the specific regions where PMSs have been reported are coronal mass ejection (CME)-driven sheaths. We use here an automated method to identify PMSs in 95 CME sheath regions observed in-situ by the Wind and ACE spacecraft between 1997 and 2015.  The occurrence and location of the PMSs are related to various shock, sheath and CME properties. We find that PMSs are ubiquitous in CME sheaths; 85\% of the studied sheath regions had PMSs with the mean duration of 6.0 hours. In about one-third of the cases the magnetic field vectors followed a single PMS plane that covered a significant part (at least 67\%) of the sheath region. Our analysis gives strong support for two suggested PMS formation mechanisms: the amplification and alignment of solar wind discontinuities near the CME-driven shock and the draping of the magnetic field lines around the CME ejecta. For example, we found that the shock and PMS plane normals generally coincided for the events where the PMSs occurred near the shock (68\% of the PMS plane normals near the shock were separated by less than $20^{\circ}$ from the shock normal), while deviations were clearly larger when PMSs occurred close to the ejecta leading edge. In addition, PMSs near the shock were generally associated with lower upstream plasma beta than the cases where PMSs occurred near the leading edge of the CME.  We also demonstrate that the planar parts of the sheath contain a higher amount of strongly southward magnetic field than the non-planar parts, suggesting that planar sheaths are more likely to drive magnetospheric activity. 
\end{abstract}


\introduction  
\label{sec:intro}
Coronal Mass Ejections (CMEs) are giant clouds of plasma and magnetic field that are expelled from the Sun into the heliosphere. CMEs often travel faster than the ambient solar wind and if their speeds exceed the local magnetosonic speed, a shock wave will develop. A turbulent region of compressed and heated plasma between the shock front and the leading edge of the CME is called a sheath region. CME-driven sheaths have an important role in solar-terrestrial studies; the parts in the immediate downstream of the shock may contribute in the acceleration of solar energetic particles \citep{manchester2005}, and when interacting with the Earth's magnetosphere, sheaths are capable of driving significant geomagnetic activity. In fact, a large number of CME-driven storms are pure-sheath induced storms (\citealp{tsurutani1988,huttunen2002,siscoe2007}). Large amplitude magnetic field and ram pressure variations within the sheath drive strong activity in particular in the high-latitude magnetosphere \citep{huttunen2004}, which may lead to large geomagnetically induced currents (GICs) \citep{huttunen2008, savani2013}. 

CME-driven  sheaths are combinations of a propagation sheath, where the solar wind flows around the obstacle in a quasi-stationary regime, and an expansion sheath, where the obstacle is expanding but not propagating with respect to the solar wind \citep{siscoe2008}. The sheath region forms gradually as the layers of interplanetary magnetic field (IMF) and solar wind plasma accumulate over several days it takes for a CME to travel from the Sun to the Earth. Hence,  CME sheaths often exhibit a complex internal structure, which makes understanding their formation and predicting their geoeffectivity particularly difficult. 

However, planar magnetic structures (PMSs) are frequently reported in CME sheaths (e.g., \citealp{nakagawa1989, neugebauer1993, savani2011}). The magnetic field vectors in a PMS are characterized by abrupt changes, but they all remain  nearly parallel to a single plane over time intervals from several hours to about one day. Such a plane includes the Parker spiral direction, but is inclined to the ecliptic plane from 30$^{\circ}$ to 85$^{\circ}$ \citep{nakagawa1989}. Hence, PMSs can be considered as laminar structures, composed of parallel planes having different orientations with respect to the interplanetary magnetic field (IMF).  During PMS events, ion densities and temperatures and plasma beta are usually higher than in the ambient plasma and the magnetic field is highly variable both in magnitude and direction \citep{nakagawa1993}. 

It has been suggested that PMSs are caused primarily by compressional processes (e.g., \citealp{jones2000, farrugia1990}). Hence, CME sheaths provide a natural environment for the formation of PMSs. Two basic mechanisms are proposed to explain how PMSs are generated in CME--sheaths. Firstly, the magnetic field lines drape around the CME ejecta and force solar wind microstructures and discontinuities to align themselves parallel to the surface of the ejecta \citep{farrugia1990}. As the magnetic field piles up at the ejecta leading edge, the compression of the IMF reduces the field variations in the direction perpendicular to the surface of the ejecta, resulting in a structure that is nearly parallel to the CME surface \citep{neugebauer1993}. The second mechanism is related to the compression of solar wind discontinuities at the CME shock, and their alignment in the downstream region parallel with the plane of the shock. In particular, PMSs are observed in the downstream regions of  quasi-perpendicular shocks \citep{jones2002} when the Alfv\'en Mach number $M_{A} > 2$ and the upstream plasma beta $0.05 < \beta < 0.5$ \citep{kataoka2005}. 

As strongly southward periods of IMF are required for efficient energy transfer from the solar wind to the magnetosphere (e.g., \citealp{dungey1961, akasofu1981}), the large amplitude out-of-ecliptic magnetic field variations related to PMSs are expected to cause significant space weather effects at the Earth. For example,  strongly southward IMF periods during the PMS-related periods in a sheath drove a significant part of the intense storm on March 17, 2015 \citep{kataoka2015} . However, it is still unclear how planarity in the sheath affects geoeffectivity, how PMSs are distributed within the sheath and what is their origin in different parts of the sheath. A better knowledge of PMS distribution and properties may help to develop early space weather forecast techniques. In addition, PMS distribution and properties may also give information on the evolution of CMEs. 

In this paper we investigate how PMSs are distributed within CME-driven sheath regions and correlate their occurrence and extension with the shock, sheath and ejecta characteristics. We also study the relationship between the orientations of the PMS plane and the CME shock normals, to check whether the plane of the shock front is driving the PMS orientation, as assumed in \cite{savani2011}, and compare the amount of significantly southward IMF in planar and non-planar cases. The structure of this paper is as follows: in Section 2 we present the data sets and the methods. In Section 3 we present statistical results on the PMS distribution and dependence on driver and sheath properties, on the relationship between PMS and shock normal orientations and on the geoeffectivity. Finally, in Section 4 we discuss and summarize our results.

\section{Data and methods}

\subsection{Data}

Our analysis involves 95 CME-driven sheath regions. The list of sheath regions is given in Table S1, Supplementary information. The investigated period covers the years from 1997 through 2015, i.e., solar cycle 23 and the rising phase and maximum of cycle 24.

The sheath region list was compiled with the help of the online UCLA interplanetary CME catalog (\url{http://www-ssc.igpp.ucla.edu/~jlan/ACE/Level3/}), which covers the time interval of 1995--2009. For the 2010--2015 period we identified interplanetary CMEs using a similar approach used to compile the UCLA catalog \citep{jian2006}. We selected only the cases where the time of the leading edge of the ejecta was well-defined, i.e. there was a clear and relatively sharp transition from the turbulent sheath to the ejecta. For discussion on the typical sheath and ejecta properties, and the difficulties in determining the CME leading edge times, see e.g., \cite{richardson2010} and  \cite{kilpua2013}, and references therein. The shock times and properties are obtained from the Heliospheric Shock Database, developed and maintained at the University of Helsinki  (\url{http://ipshocks.fi/}). 

Solar wind and IMF measurements are taken primarily from the Wind spacecraft. Wind was launched in November 1994. It spent its first years at the Lagrangian point L1 and in 1999 it acquired a complex petal-shaped orbit that brought it further away from the Sun-Earth line. Since 2004, Wind has been operating at L1. The data from the Advanced Composition Explorer (ACE) spacecraft is used during Wind's magnetospheric visits and datagaps. ACE was launched in August 1997 and it has been operating close to L1. 

We use data from the Wind Magnetic Fields Investigation (MFI) \citep{lepping1995} and the Wind Solar Wind Experiment (SWE) \citep{ogilvie1995}. MFI data are available at 60-second resolution and SWE data are registered about every 90 seconds. From ACE, we use Magnetic Field Investigation (MFI) \citep{smith1998} data, available from September 1997, and ACE Solar Wind Electron Proton Alpha Monitor (SWEPAM) \citep{mccomas1998} data, available from February 1998. MFI and SWEPAM data are 16-second and 64-second level 2 data,  respectively. Both Wind and ACE data are obtained from the NASA Goddard Space Flight Center Coordinated Data Analysis Web (CDAWeb, \url{http://cdaweb.gsfc.nasa.gov/}).

\subsection{PMS identification method}
\label{subsec:method}

\begin{figure}[t]
\vspace*{2mm}
\begin{center}
\includegraphics[width=8.3cm]{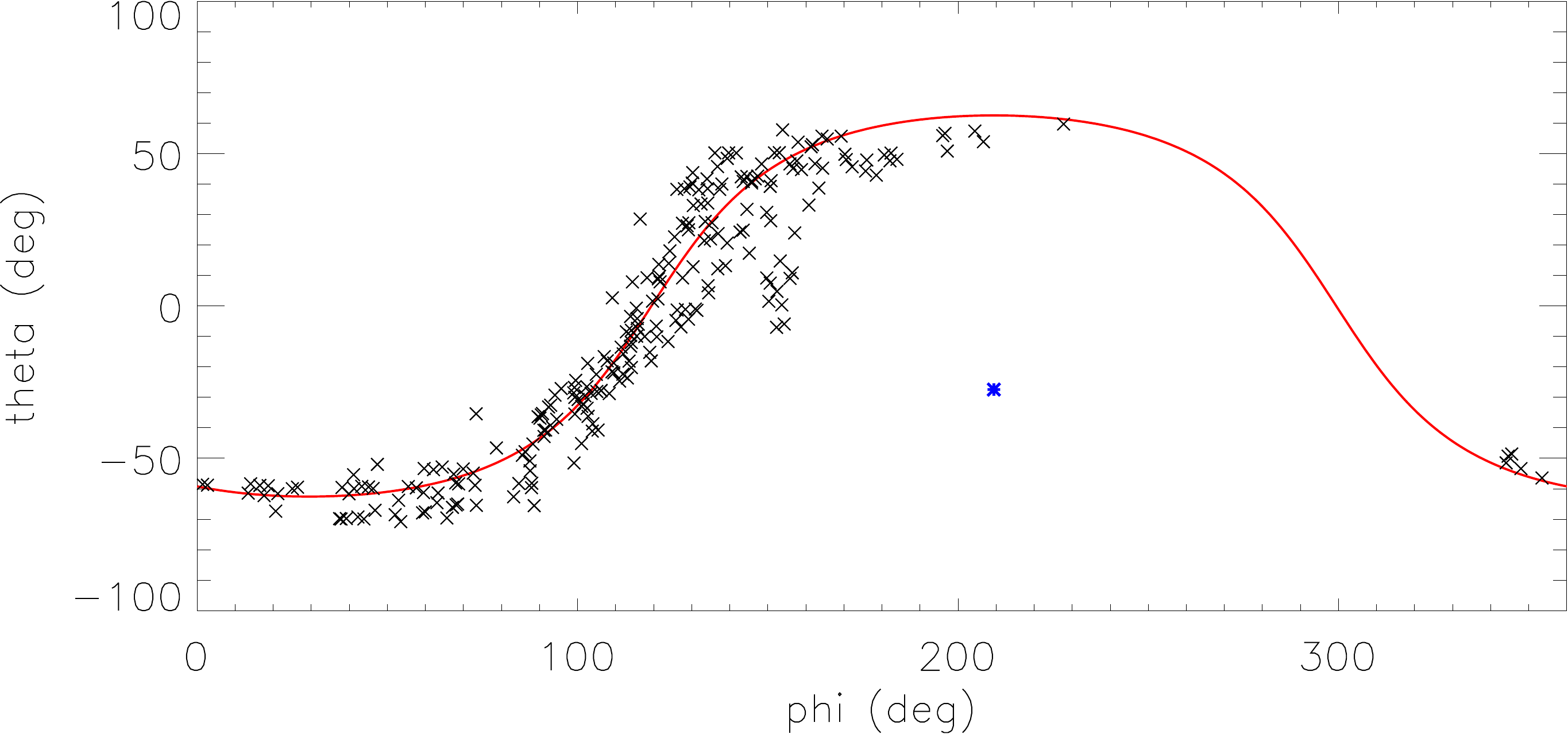}
\end{center}
\caption{An example of a $\theta$-$\phi$ diagram. The direction normal to the PMS plane is displayed by the blue ``$\ast$'' and the plane perpendicular to it is represented by the red curve. The black scatter points are the magnetic field vectors as observed by Wind within the sheath region following an interplanetary shock on January 10, 1997.} 
\label{fig:pmscurve}
\end{figure}

We describe next our approach to identify PMSs within the sheath regions. As discussed in Introduction, a PMS is defined as an interval where the IMF vectors remain parallel to a fixed plane for an extended time period (of the order of hours). Let's assume that the IMF field vectors in the Geocentric Solar Ecliptic (GSE) coordinates are $\mathbf{B} \equiv (B_{X}, B_{Y}, B_{Z}) \equiv (B\cos\theta\cos\phi, B\cos\theta\sin\phi, B\sin\theta)$ ($\phi$ and $\theta$ are the IMF  longitude and latitude, respectively) and that they are all parallel to a fixed plane with the normal $\mathbf{n} \equiv (n_{X}, n_{Y}, n_{Z})$. Now the relation between $\theta$ and $\phi$ is 
\begin{equation}
n_{x}\cos\theta\cos\phi + n_{y}\cos\theta\sin\phi + n_{z}\sin\theta = 0 \, .
\label{eq:pmscurve}  
\end{equation}
Hence, PMSs are identified as the periods where  in the $\theta$-$\phi$ space the IMF vectors are distributed in the close proximity of the curve presented by Eq. (\ref{eq:pmscurve}), see example from Fig. \ref{fig:pmscurve}. Moreover, in a PMS event IMF data points in a $\theta$-$\phi$ diagram cover a wide range of $\phi$ values \citep{nakagawa1989}, meaning that they can assume almost any direction within the plane.

To determine the orientation of the PMS plane we use the minimum variance analysis (MVA) \citep{sonnerup1998}. The minimum variance direction of the IMF vectors corresponds to the normal of the PMS plane. The ratio of the intermediate ($\lambda_{2}$) to minimum ($\lambda_{3}$) eigenvalue can be used as a proxy of the quality of the MVA and we use here the consistency requirement $\lambda_{2}/\lambda_{3} \geq 5$  (e.g., \citealp{savani2011}). The quality of the result increases with the increasing eigenvalue ratio and the solution is degenerate when $\lambda_{2}\simeq\lambda_{3}$. The two-dimensionality of the field vectors in PMS events can be confirmed through the value of $|B_{n}|/B$ \citep{nakagawa1993}, where $B$ is the magnetic field magnitude and $B_{n} = \mathbf{B} \cdot \mathbf{n}$ is the component of the magnetic field perpendicular to the PMS plane. We impose a requirement that $|B_{n}|/B < 0.2$  (e.g., \citealp{jones2000} and \citealp{kataoka2005}). 

We apply the MVA and detect PMS candidates through an automated procedure. The aim of the method is to scan, within the sheath, all the possible combinations of PMS intervals with 5 minutes time steps, starting from the largest duration (full sheath) to the minimum duration (1 hour). This algorithm is more effective than a simple sliding window, as two adjacent intervals that satisfy the MVA criteria may actually correspond to two separate PMSs with different planes. Hence, our method avoids mixing two different PMS intervals. The algorithm is implemented as follows. First, the MVA is performed over the whole sheath and the eigenvalue ratio and $|B_{n}|/B$ are stored for final comparison. Thereafter a five-minute interval is cut from the leading edge (LE) side, a new MVA is applied and its result is again recorded. This reduced window is then shifted five minutes towards the LE (at the first time interfacing the edge) and MVA is applied again. Then the initial full-sheath window is shortened by additional five minutes from the LE side, MVA is applied, and thereafter the 10-min reduced window is shifted again by subsequent five-minute steps until it reaches again the LE interface. The entire procedure is repeated, decreasing each time the initial full-sheath window length by additional five minutes until the minimum duration of one hour is reached. At each step the eigenvalue ratio and $|B_{n}|/B$ are recorded and finally compared. The PMS intervals are the longest non-overlapping intervals fulfilling the requirements described above. In a case of multiple windows of the same length satisfying our criteria, we select the interval which features the lowest $|B_{n}|/B$. Note that if the entire sheath would be one single PMS, already the first MVA would give the best result. We also require for the selection that  the longitudinal coverage of the magnetic field vectors along the PMS curve has to exceed 90$^{\circ}$ \citep{jones2000}.

\begin{figure}[t]
\vspace*{2mm}
\begin{center}
\includegraphics[width=8.3cm]{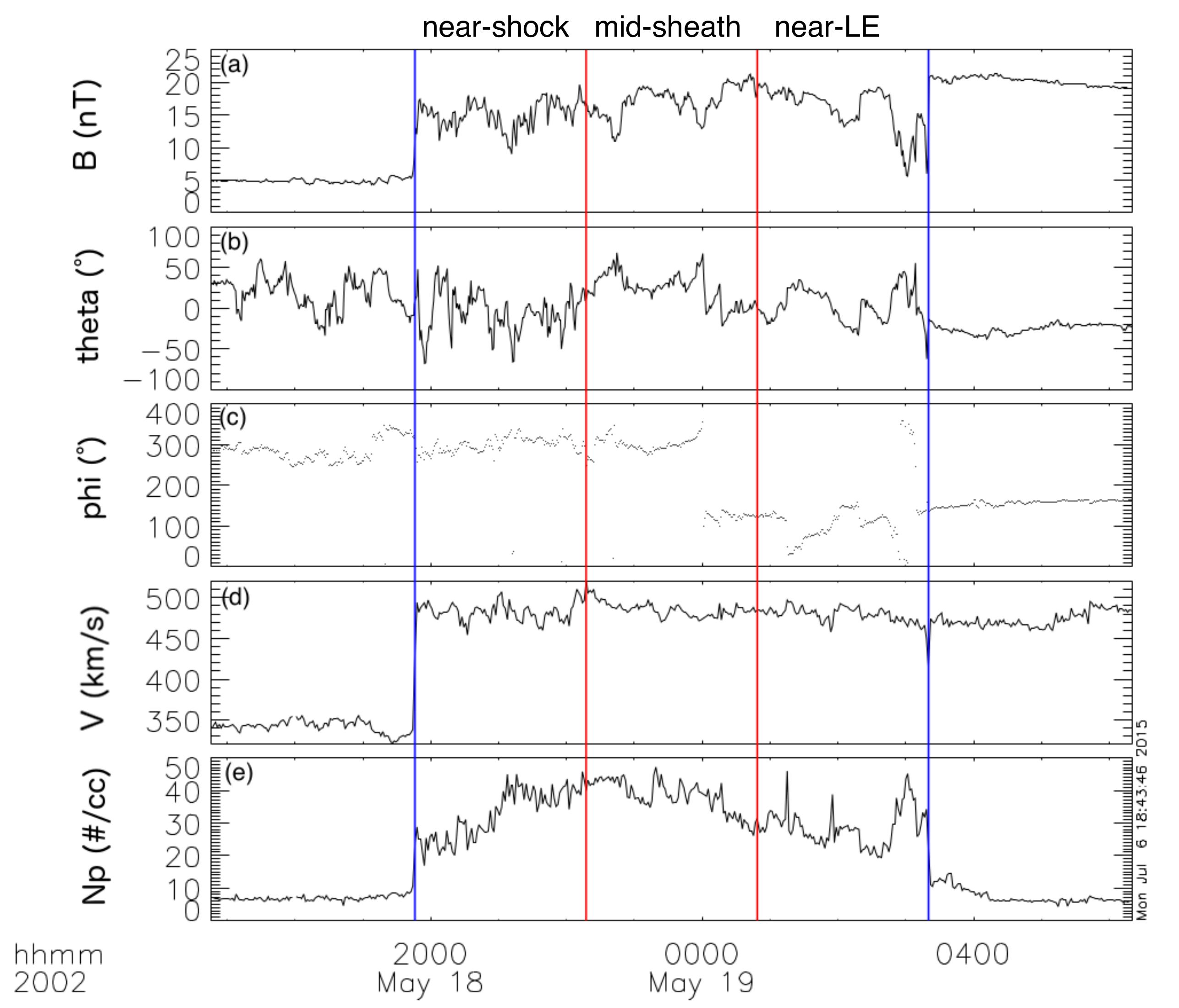}
\end{center}
\caption{A sheath region on May 18, 2002 observed by Wind. The blue lines indicate the CME shock and the ejecta leading edge (LE), respectively, while the red lines separate the sheath into 33\%-33\%-33\% sub-regions: Near-Shock, Mid-Sheath and Near-LE domains.  The shown parameters are from top to bottom: (a) magnetic field 
magnitude, (b) $\theta$ and (c) $\phi$ components in GSE angular coordinates, 
(d) solar wind speed and (e) proton density.} 
\label{fig:regions}
\end{figure}

To investigate the distribution of PMSs we divide sheaths into three sub-regions (Fig. \ref{fig:regions}): Near-Shock, Mid-Sheath and Near-Leading-Edge (Near-LE) regions.  Then the sheath regions are placed into five groups according to the following criteria:
\begin{itemize}
\item \textbf{Group 1 -- no PMS:}  No PMSs are found (0\% PMS coverage in all three sub-regions)
\item \textbf{Group 2 -- Near-Shock PMS:}  PMSs starts at the vicinity of the IP shock (latest one hour from the shock) and they cover at least 67\% of the Near-Shock region. PMSs may extend in the Mid-Sheath region
\item \textbf{Group 3 -- Mid-Sheath PMS:} PMSs cover at least 67\% of the Mid-Sheath region, but less than  67\% of the Near-Shock and Near-LE regions.
\item \textbf{Group 4 -- Near-LE PMS:}  PMSs starts at the vicinity of the CME leading edge (latest one hour from the leading edge) and they cover at least 67\% of the Near-LE region. PMSs may extend in the Mid-Sheath region
\item \textbf{Group 5 -- Full-PMS:}  PMSs cover at least 67\% of all three regions 
\end{itemize}
From the total of 95 analyzed events, eight sheaths were not included in any of the above described groups, i.e. at least one PMS was found, but its coverage was $<$ 67\% in all three regions.  

\section{Results}

\subsection{PMS coverage}

\begin{figure}[t]
\vspace*{2mm}
\begin{center}
\includegraphics[width=8.3cm]{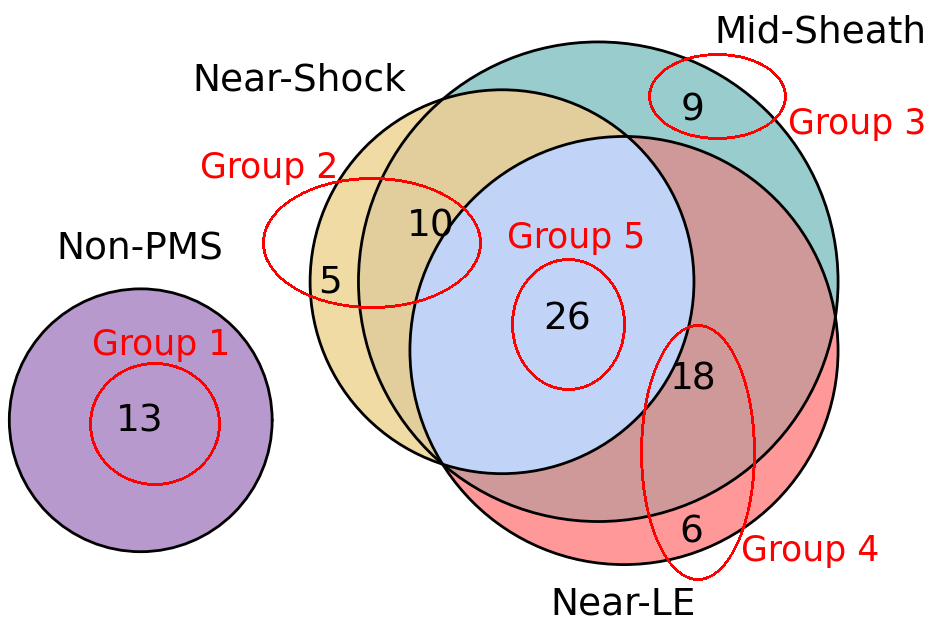}
\end{center}
\caption{Distribution of the investigated sheath regions according to the presence and position of PMSs (for the explanation of different groups see Section \ref{subsec:method}).} 
\label{fig:venn}
\end{figure}

Figure \ref{fig:venn} shows that 86\% of the investigated sheaths  (82 of total 95) present at least one PMS. The 67\%-coverage criterion was satisfied for 43\% of the Near-Shock regions (41 events), 66\% of the  Mid-Sheath regions (63 events) and  53\% (50  events) of the Near-LE regions. The distribution of the events in five PMS groups (see Section 2.2) is also seen from Fig. \ref{fig:venn}: for 13 sheaths no PMSs were found (Group 1), 15 sheaths are in the Near-Shock PMS group (Group 2), nine in the Mid-Sheath group (Group 3), 24 in the Near-LE group (Group 4), and 26 events are the cases where the 67\%-coverage criteria was satisfied in all three sub-regions, i.e. ``full PMS'' cases (Group 5). In addition, as mentioned in Section 2.2, there were eight events featuring PMS, but for which the 67\%-coverage criterion was not satisfied in any of the three sub-regions. The durations of the identified PMS events range from 1.3 to 20.5 hours, with an average duration of 6.0 hours. 

In the majority of the cases featuring PMSs (74 of total 82) we found only one PMS plane. Seven sheath regions featured two distinct PMS intervals with different plane orientations. Only one event (July 17, 2005) had three PMS periods with different plane orientations. The sheaths with two different PMS planes are distributed in the following way: three in Group 2, two in Group 3, one in Group 4 and one in Group 5. The event with three different PMS planes does not belong to any of our groups, since PMS durations did not meet our 67\%-coverage criterion.  
 
\subsection{Dependence on shock, sheath and CME ejecta properties}

\begin{figure*}[!ht]
  \centering
  \subfloat{\includegraphics[width=.26\textwidth]{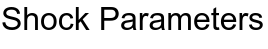}}\\
  \addtocounter{subfigure}{-1}
  \subfloat[][]{\includegraphics[width=.235\textwidth]{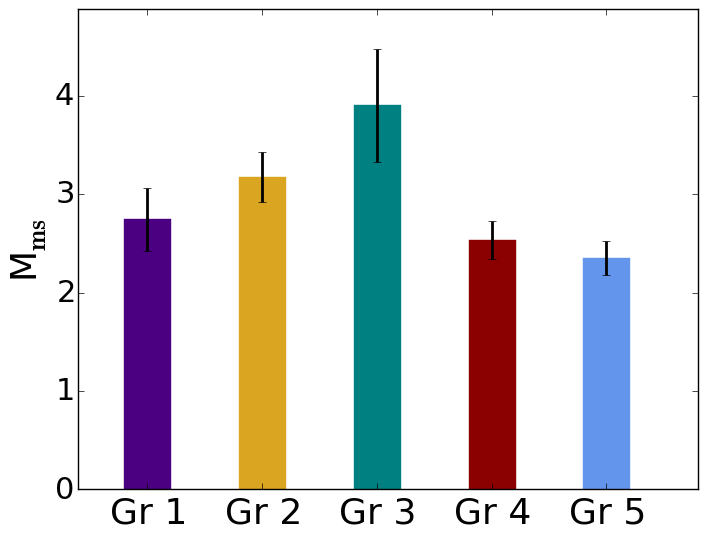}}\quad  
  \subfloat[][]{\includegraphics[width=.235\textwidth]{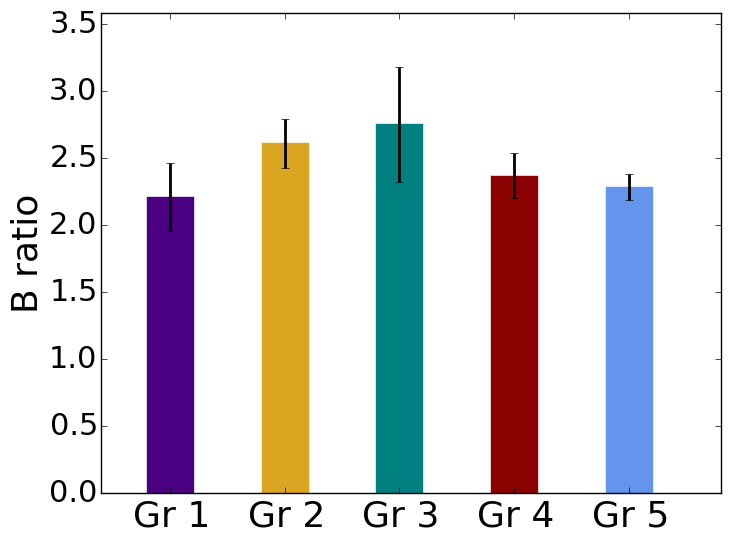}}\quad
  \subfloat[][]{\includegraphics[width=.235\textwidth]{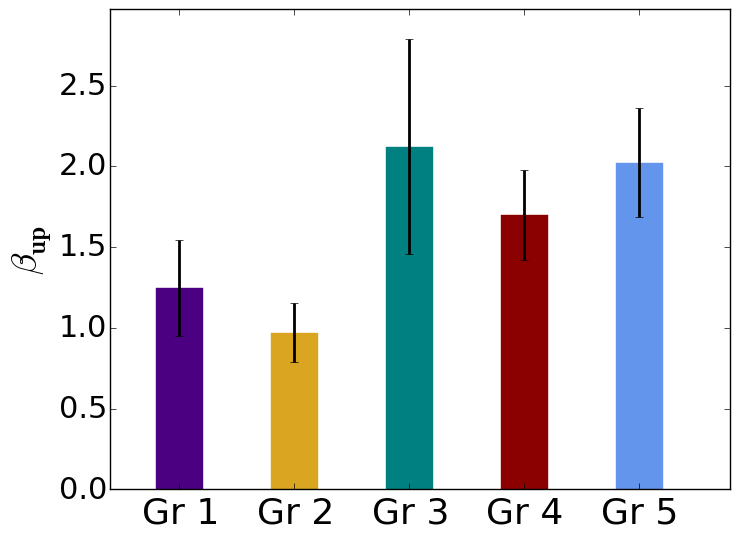}}\quad
  \subfloat[][]{\includegraphics[width=.235\textwidth]{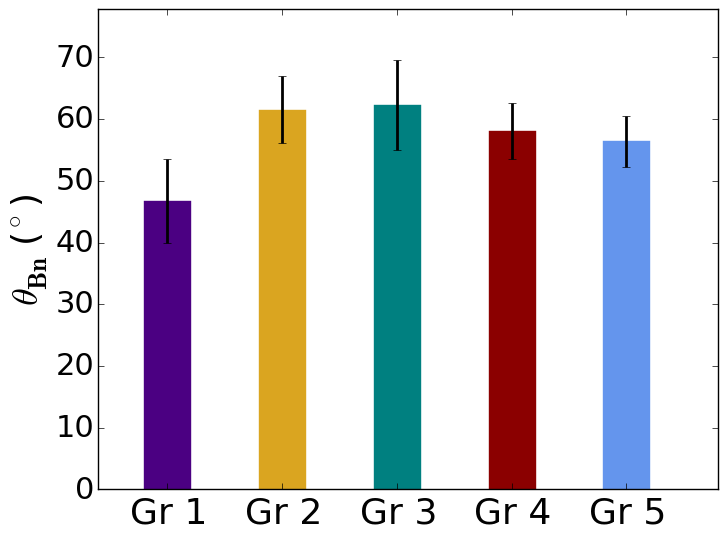}}\\
  \subfloat{\includegraphics[width=.27\textwidth]{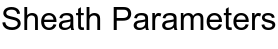}}\\
  \addtocounter{subfigure}{-1}
  \subfloat[][]{\includegraphics[width=.235\textwidth]{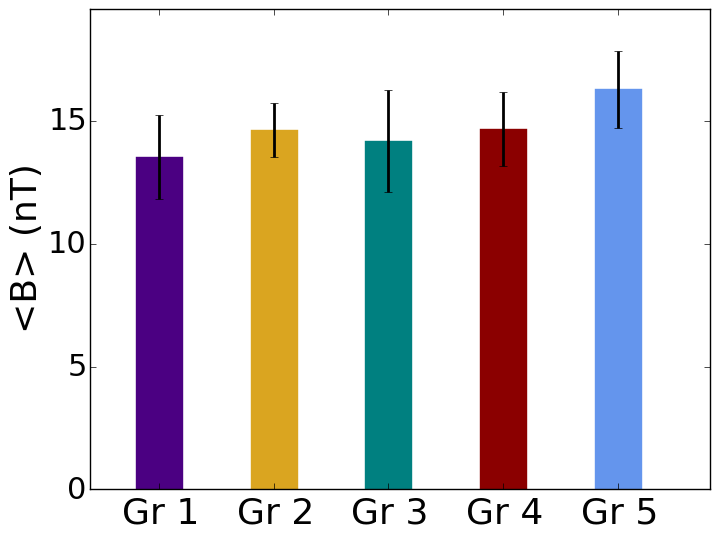}}\quad
  \subfloat[][]{\includegraphics[width=.235\textwidth]{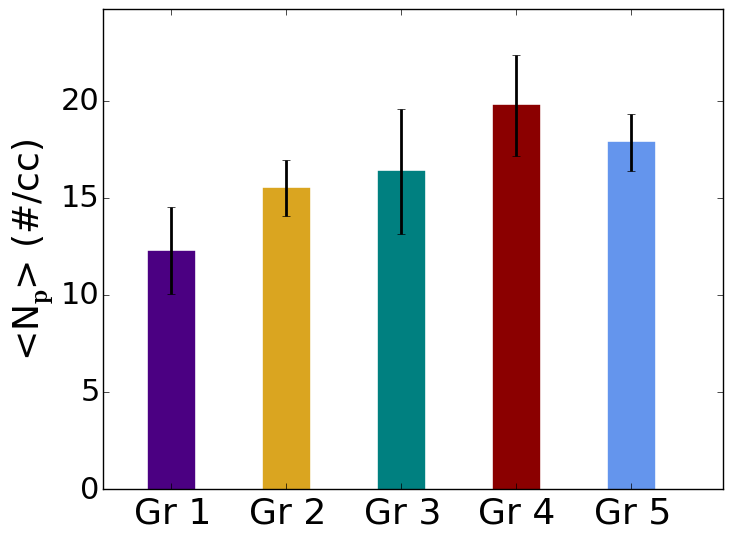}}\quad
  \subfloat[][]{\includegraphics[width=.235\textwidth]{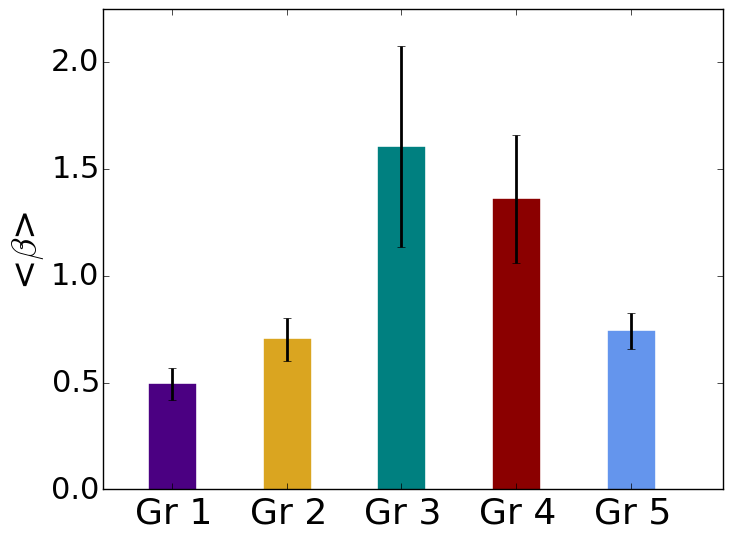}}\quad
  \subfloat[][]{\includegraphics[width=.235\textwidth]{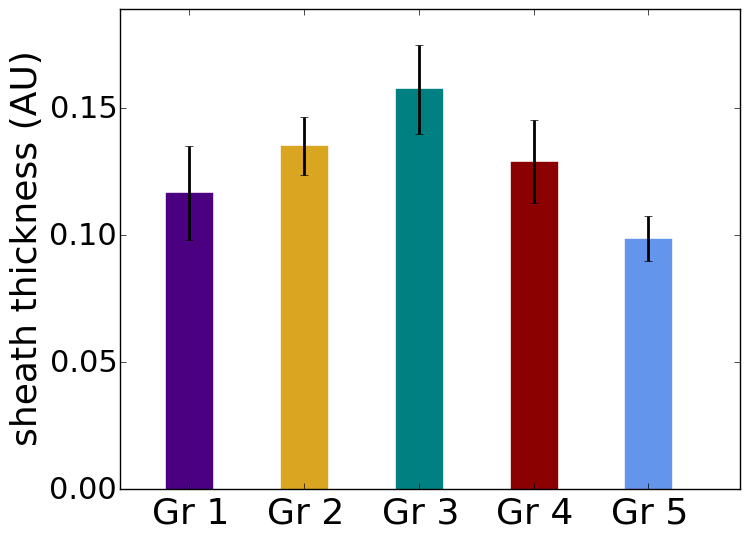}}\\
  \subfloat{\includegraphics[width=.26\textwidth]{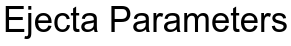}}\\
  \addtocounter{subfigure}{-1}
  \subfloat[][]{\includegraphics[width=.235\textwidth]{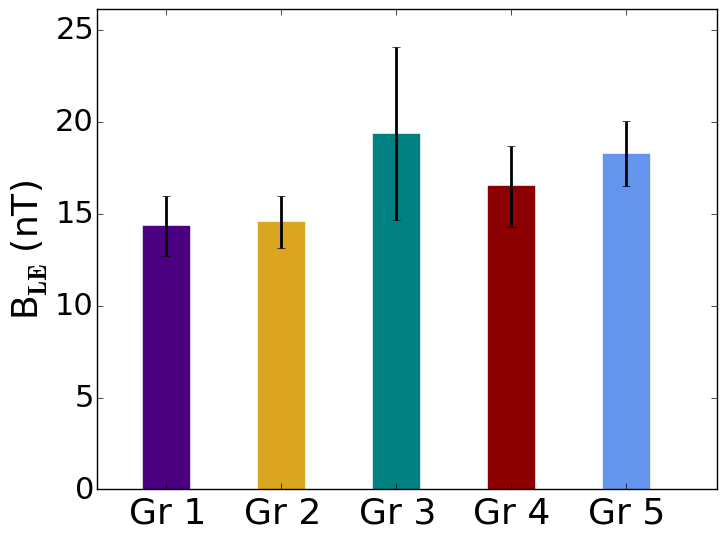}}\quad
  \subfloat[][]{\includegraphics[width=.235\textwidth]{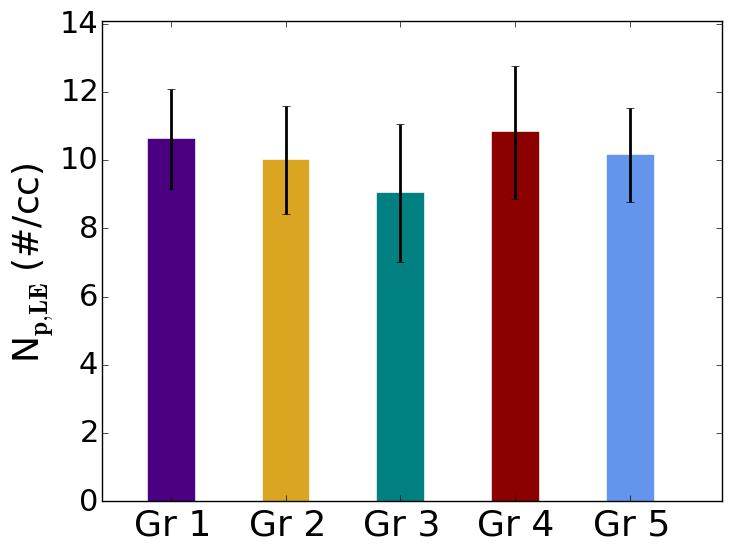}}\quad
  \subfloat[][]{\includegraphics[width=.235\textwidth]{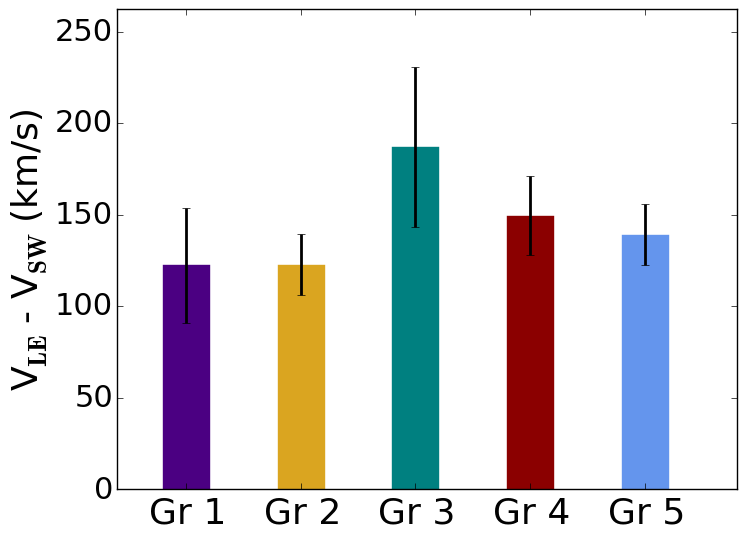}}\quad
  \subfloat[][]{\includegraphics[width=.235\textwidth]{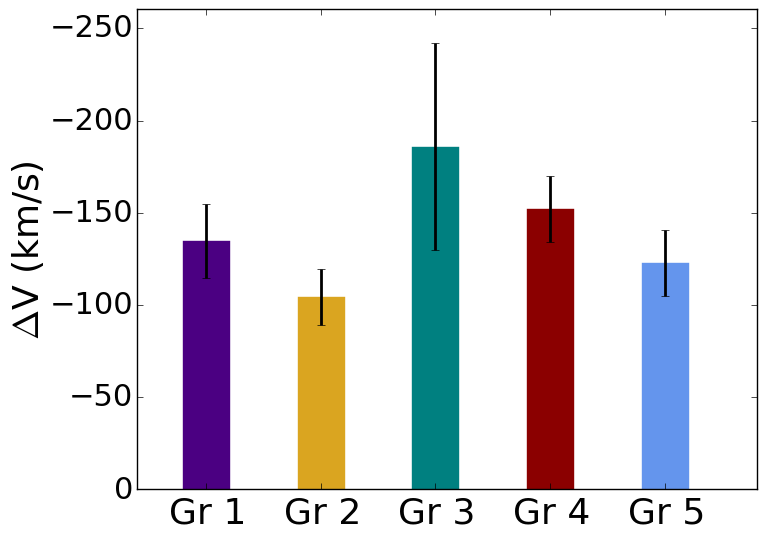}}
  \caption{The investigated parameters and their relation to PMS groups. Histograms show the averages, and the error bars show  the extent of one standard deviation. First row -- shock parameters: (a) magnetosonic Mach number, (b) downstream-to-upstream magnetic field magnitude ratio, (c) upstream plasma beta, (d) shock normal. Second row -- sheath parameters: (e) average magnetic field magnitude, (f) average proton number density, (g) average plasma beta, (h) sheath thickness. Third row -- CME ejecta parameters: (i) leading edge magnetic field magnitude, (j) leading edge proton number density, (k) difference between leading edge and upstream speed, (l) velocity gradient (only for expanding CME ejecta).}
  \label{fig:histograms}
\end{figure*}

Figure \ref{fig:histograms} shows the results of our statistical analysis on the relation between the PMS distribution and the shock, sheath and ejecta properties. 

\subsubsection{Shock parameters}

The averages of selected shock parameters are presented in the top row of Fig. \ref{fig:histograms}. The magnetosonic Mach number $M_{ms}$ (panel \ref{fig:histograms}a) is an indicator of the strength of the shock (e.g., \citealp{burgess1995}), i.e. the estimate of the energy processed by the shock. It should be noted that $M_{ms}$ depends on the shock angle and the magnetosonic speed, which may have large errors. Hence, we show the results also for the downstream-to-upstream magnetic field magnitude ratio (panel \ref{fig:histograms}b), which typically increases with increasing $M_{ms}$  (e.g., \citealp{burton1996}). Panels (a) and (b) of Fig. \ref{fig:histograms} indeed show very similar characteristics. The highest values of $M_{ms}$ and the magnetic field ratios are attributed to Group 2 (Near-Shock) and Group 3 (Mid-Sheath). 

The upstream plasma beta $\beta_{up}$ (panel \ref{fig:histograms}c), i.e., the ratio between plasma and magnetic pressure, describes the upstream plasma state. The histogram shows that the lowest values of $\beta_{up}$ are found for Group 2 (Near-Shock), where $\beta_{up} \approx 1$, while the highest values belong to Group 3 (Mid-Sheath) and Group 5 (Full-PMS), with $\beta_{up} \approx 2$.

The shock angle $\theta_{Bn}$ (panel \ref{fig:histograms}d) between the upstream magnetic field direction and the shock normal describes the nature and structure of the IP shock. At quasi-parallel shocks (i.e. $\theta_{Bn} < 45^{\circ}$) the magnetic field lines crossing the shock carry particles through it relatively easily, resulting in broad and gradual transitions in the plasma and magnetic field parameters across the shock (e.g., \citealp{burgess2005}). In contrast, at quasi-perpendicular shocks (i.e. $\theta_{Bn} > 45^{\circ}$), the magnetic field lines are more parallel to the surface of the shock and the particles gyrating along the field lines reflect back to the shock, resulting in a steep change in plasma parameters and in the magnetic field magnitudes (e.g., \citealp{bale2005}). Panel \ref{fig:histograms}d shows that non-planar sheaths (Group 1) are associated with the most parallel shocks when compared to planar sheaths. The most perpendicular shocks are found for Group 2 (Near-Shock) and Group 3 (Mid-Sheath), with a mean value $\theta_{Bn} \approx 60^{\circ}$. 

\subsubsection{Sheath parameters}

The mid row of Fig. \ref{fig:histograms} shows the average magnetic field magnitude, plasma density and plasma $\beta$ within the sheath as well as the sheath thickness (i.e., the average sheath duration multiplied by the average speed within the sheath). The magnetic field magnitudes (panel \ref{fig:histograms}e) are rather similar for all investigated groups. The lowest value is found for non-planar cases and the highest value for full-PMS cases. The non-planar sheaths have also the lowest average densities and plasma $\beta$. For planar sheaths the lowest density and plasma $\beta$  are found for the Near-Shock group (Group 2). The average plasma $\beta$ is also low for the Full-PMS group (Group 5). Group 3 (Mid-Sheath) features the highest plasma $\beta$, but the corresponding error bar is quite large. Panel \ref{fig:histograms}h shows that Full-PMS sheaths are shortest  ($\sim 0.09$ AU) when compared with the other groups, which have a mean thickness of $\sim 0.13$ AU. 

\subsubsection{Ejecta parameters}

The first two panels in the bottom row of  Fig.  \ref{fig:histograms} show the CME leading edge magnetic field magnitude and proton number density, respectively, calculated here as three-hour averages from the leading edge time onwards through the CME ejecta. The non-planar sheaths and the events in the Near-Shock group are associated with weaker leading edge magnetic field magnitudes than the other groups. The comparison of panels \ref{fig:histograms}f and \ref{fig:histograms}j shows that non-planar cases (Group 1) have very similar densities both at the ejecta leading edge and in the sheath, while for planar events (Groups 2-5) the densities are clearly larger in the sheath and then drop to values comparable with non-PMS events at the ejecta leading edge.

Panel \ref{fig:histograms}k shows the speed difference between the leading edge and the solar wind in front of the sheath. The CME leading edge speed is calculated as the three-hour average from the leading edge time through the ejecta, while the upstream speed is calculated as the solar wind velocity average during the three hours immediately before the IP shock. At CME flanks the shock is essentially a blast-wave type and its speed decreases with distance \citep{pick2002}. In total there were four events (4\%) with negative speed differences that were not included in the analysis. The highest speed difference is found for group 3 (Mid-Sheath), while the smallest differences are associated with Group 1 (Non-PMS) and Group 2 (Near-Shock).

The last panel in the bottom row shows the averages of the CME velocity gradient, defined as the difference in the velocities between the CME trailing and leading edges. Only cases featuring negative velocity gradients, i.e. expanding ejecta, are taken into account in  Fig. \ref{fig:histograms}l.  This histogram shows a similar trend to the previous one (panel \ref{fig:histograms}k) and the strongest expansion is found for events in Group 3 (Mid-Sheath) and Group 4 (Near-LE), while the events in Group 2 (Near-Shock) have on average the weakest expansion. In total there were five events (6\%) with positive velocity gradients, i.e. the CME ejecta was compressed in such cases. All these events are full-PMS cases. Two of the ejecta had nearly zero velocity gradients. The associated sheaths belong to Group 3 (Mid-Sheath) and to Group 5 (Full-PMS).

\subsection{Plasma parameters in the sheath sub-regions}

\begin{figure*}[!ht]
  \centering
  \subfloat[][]{\includegraphics[width=.25\textwidth]{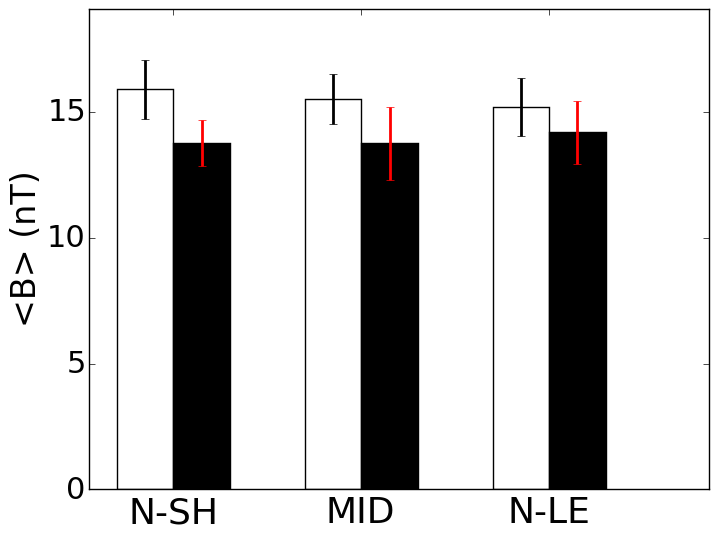}}\quad  
  \subfloat[][]{\includegraphics[width=.25\textwidth]{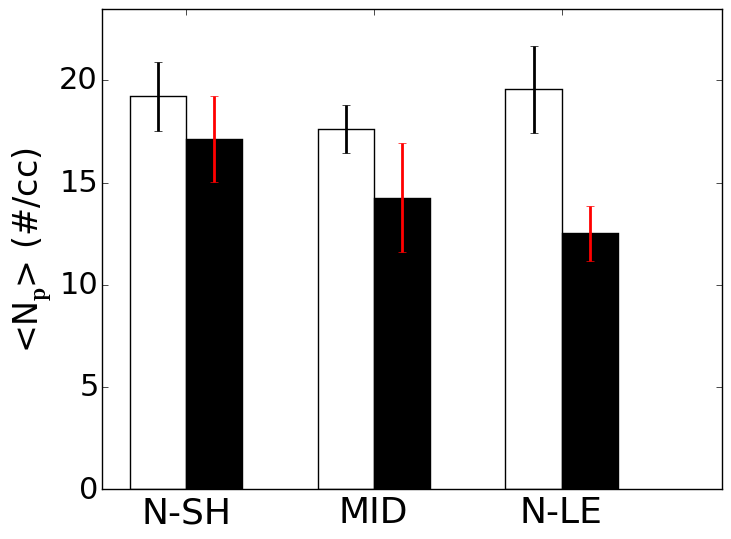}}\quad
  \subfloat[][]{\includegraphics[width=.25\textwidth]{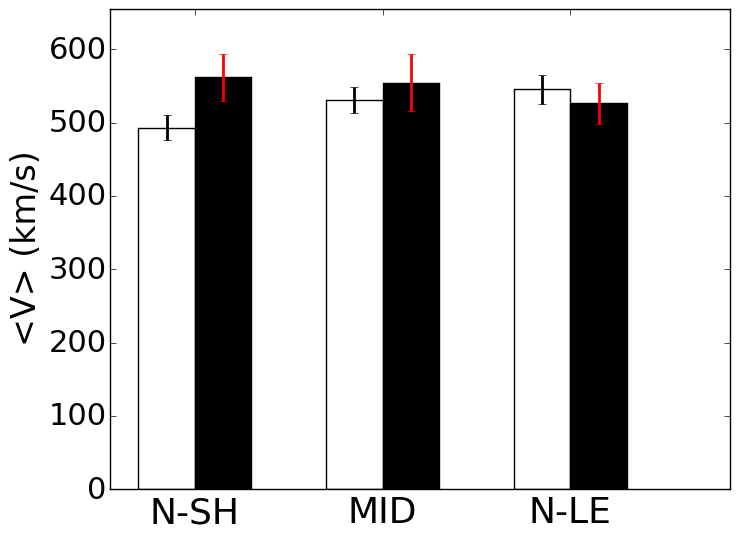}}\\
  \subfloat[][]{\includegraphics[width=.25\textwidth]{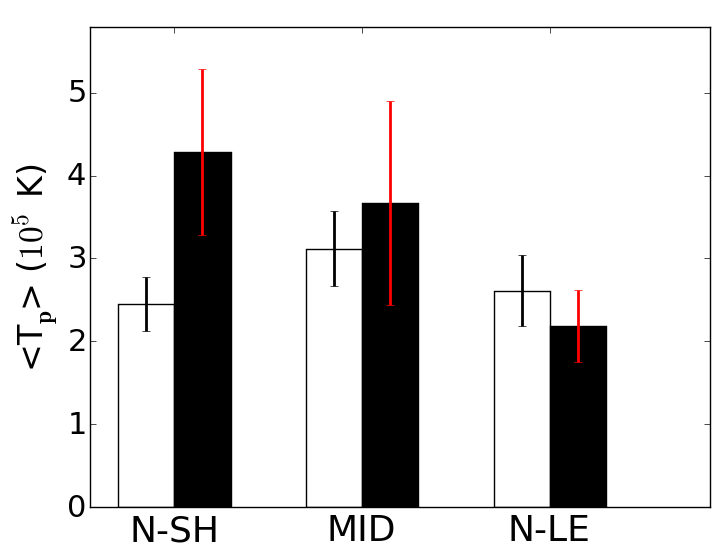}}\quad
  \subfloat[][]{\includegraphics[width=.25\textwidth]{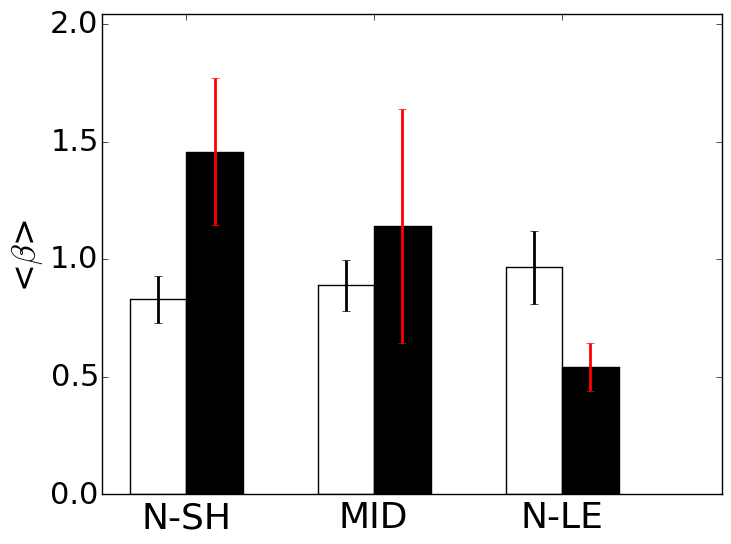}}\quad
  \subfloat[][]{\includegraphics[width=.25\textwidth]{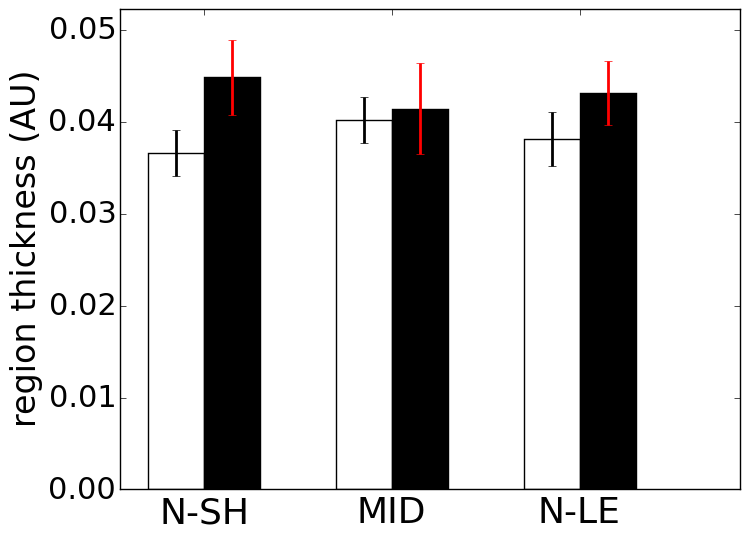}}
  \caption{Dependence of PMS occurrence on average sheath plasma parameters in each sheath sub-regions (Near-Shock: N-SH, Mid-Sheath: MID and Near-LE: N-LE).  The white bars represent the averages within the sub-regions where our PMS criteria were satisfied (67\% PMS coverage), while the black bars show the averages in non-planar sub-regions (0\% PMS coverage).  The black and red error bars show the extent of one standard deviation. The studied parameters are: (a) magnetic field magnitude, (b) proton number density, (c) solar wind speed, (d) proton temperature, (e) plasma beta, and (f) sub-region thickness.}
  \label{fig:histograms2}
\end{figure*}

As discussed in Introduction, previous studies suggest that local solar wind conditions have a significant role in the formation of PMSs. While Fig. \ref{fig:histograms} showed selected solar wind parameters averaged over the whole sheath, in Fig. \ref{fig:histograms2} we present the averages of a more complete set of plasma parameters in three sheath sub-regions (Near-Shock, Mid-Sheath, Near-LE regions, see Fig. \ref{fig:regions}). The white histograms show the cases where our PMS criterion was fulfilled (in 41 Near-Shock regions, 63 Mid-Sheath regions and 50 Near-LE regions) and the black histograms the cases where no PMS was identified (in 30 Near-Shock regions, 19 Mid-Sheath regions and 29 Near-LE regions). 

Figure \ref{fig:histograms2} shows that the mean magnetic field magnitude (panel \ref{fig:histograms2}a), solar wind speed (panel \ref{fig:histograms2}c), and the thickness of the sub-region (panel \ref{fig:histograms2}f) do not vary significantly between non-planar and planar parts of the sheath region.

In turn, plasma $\beta$, solar wind density and temperature show clear variations. Those regions in which the 67\% PMS-coverage PMS criteria is fulfilled have higher density (panel \ref{fig:histograms2}b) than their non-planar counterparts. This trend is most pronounced in the Near-LE region. For the Near-Shock region temperatures (panel \ref{fig:histograms2}d)  are clearly lower when PMSs are present, while for PMS-related Near-LE regions the average temperature is  even slightly higher than in a case of non-planarity.  It is interesting to note that average temperatures are rather similar in all three sub-regions when PMSs were identified, while the averages show large variations for non-planar sub-regions. The same trend is also visible for the plasma $\beta$ (panel \ref{fig:histograms2}e). In the Mid-Sheath region, and in particular in the Near-Shock region, plasma $\beta$ is higher for non-planar cases, while in the Near-LE region $\beta$ is higher when a significant PMSs coverage is found.  

\subsection{Relation between PMS and shock normal orientations}

\begin{figure}[t]
\vspace*{2mm}
\begin{center}
\includegraphics[width=8.3cm]{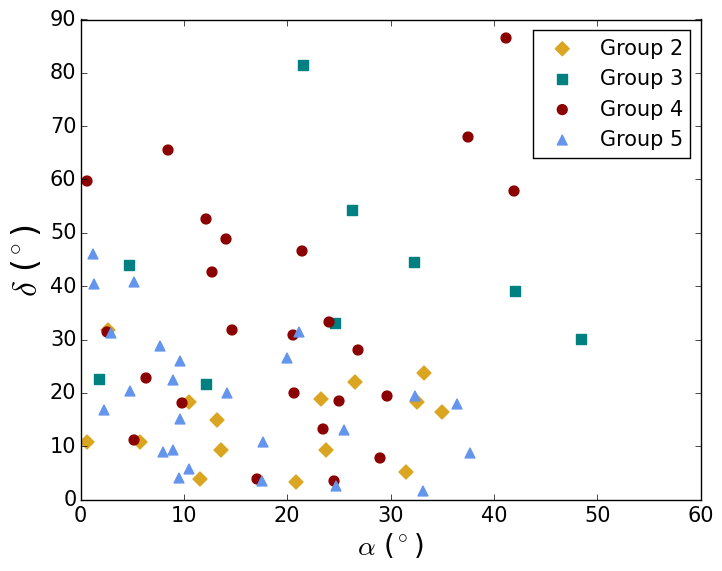}
\end{center}
\caption{Distribution of the angle $\delta$ (between PMS normal and shock normal) versus the angle $\alpha$ (between shock angle and radial distance, absolute value). The different colors and symbols refer to the PMS groups: Near-Shock PMS (yellow diamonds), Mid-Sheath PMS (green squares), Near-LE PMS (red circles), Full-PMS sheath (blue triangles).} 
\label{fig:standoff}
\end{figure}

Figure \ref{fig:standoff} shows how the angle between the PMS normal and the shock normal ($\delta$) are related to the angle that the shock normal makes with the radial direction ($\alpha$), i.e., when $\delta=0^{\circ}$ the shock normal is aligned with the PMS normal. The radial direction corresponds to the Sun-Earth direction ($-\mathbf{\hat{x}}_{GSE}$). The location angle $\alpha$ is used here as an approximation of the spacecraft crossing distance from the CME apex (e.g., \citealp{janvier2015, savani2015}), where for values close to zero the spacecraft is crossing the ejecta close to its apex and the angle increases as the crossing takes place more on the flank of the CME.

It is clear from Fig. \ref{fig:standoff} that there is no correlation between $\delta$ and $\alpha$ (correlation coefficient $R = 0.08$). In turn, the color-coded scatter points present a clear dependence of the PMS location on the value of $\delta$. For Group 2 (Near-Shock) the normal of the  PMS plane is more or less aligned with the shock normal ($\delta$ does not exceed $\approx 32^{\circ}$ for the yellow points). Group 5 (Full-PMS), that also features a PMS initiating within one hour after the IP shock, shows similar characteristics ($\delta$ does not exceed $\approx 48^{\circ}$ for the blue points). In turn, Group 3 and Group 4 have a significantly larger range of $\delta$ values (reaching nearly $90^{\circ}$) and a tendency for PMS plane having considerable deviation from the shock normal.

\subsection{Relationship between PMSs and southward magnetic field}

\begin{figure*}[!ht]
  \centering
  \subfloat[][]{\includegraphics[width=.45\textwidth]{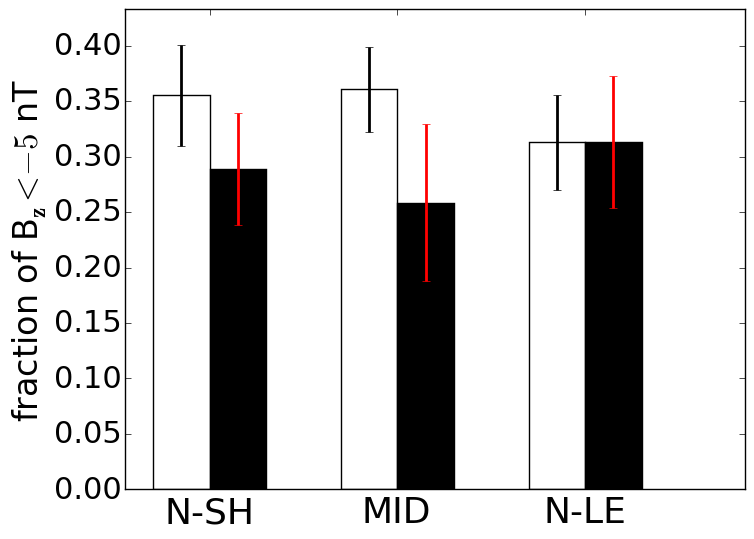}}\quad  
  \subfloat[][]{\includegraphics[width=.45\textwidth]{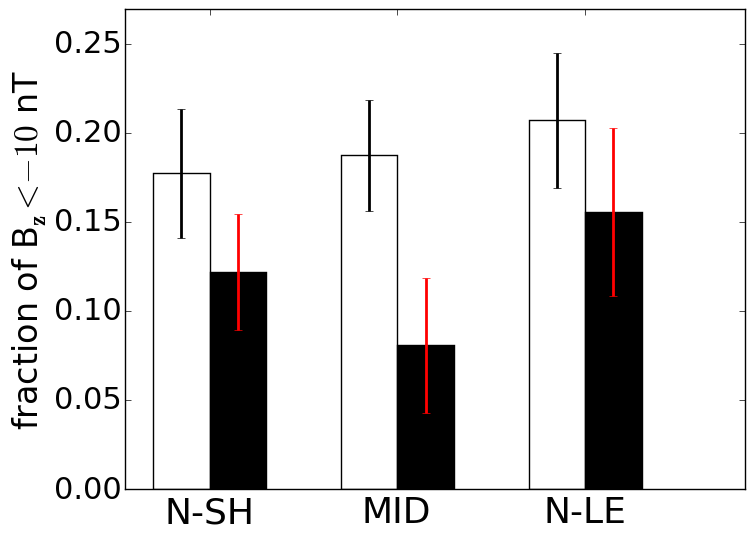}}\\
  \caption{Dependence of PMS occurrence on southward magnetic field $B_{z}$ (in GSM coordinates) in the same format as Figure \ref{fig:histograms2}. The investigated parameter is the fraction of sub-region covered by (a)  $B_{z} < -5$ nT, (b) $B_{z} < -10$ nT. The error bars show the extent of one standard deviation.} 
\label{fig:geo}
\end{figure*}

Finally, we investigate whether the presence of PMSs could influence the geoeffectivity of the sheath region.  Figure \ref{fig:geo} presents the fractions of duration covered by $B_{z} < -5$ nT and $B_{z} < -10$  nT in the three sheath sub-regions (Near-Shock, Mid-Sheath, Near-LE regions, see Figure \ref{fig:regions}). The results are presented in the same format as Fig. \ref{fig:histograms2}: the white histograms show the cases where our 67\%-coverage PMS criteria was fulfilled and the black histograms the cases where no PMSs were identified (0\% PMS occurrence). 

Both panels suggest that planar regions are more likely to have strongly southward magnetic fields than non-planar regions. The Near-LE region has equal fraction of $B_{z} < -5$ nT for PMS and non-PMS cases, but there is a higher fraction of intense southward magnetic field ($B_{z} < -10$ nT) in PMS-related Near-LE regions. In particular, the Mid-Sheath region has significantly more $B_{z} < -10$ nT intervals when PMSs are found.

\section{Discussion and conclusions}  
In this paper we have studied the occurrence and distribution of planar magnetic structures (PMSs) within 95 CME-driven sheath regions and their association with the shock, sheath and ICME properties. The sheath regions under study were divided in three sub-regions (Near-Shock, Mid-Sheath, and Near-LE; LE=leading edge) and subsequently in five groups according to PMS presence and location within the sheath (see Fig. \ref{fig:venn} and Sect. \ref{subsec:method}).

Our study shows that PMSs are ubiquitous in CME sheaths; 85\% of the analyzed sheaths feature at least one PMS and in over one-third (35\%) of the cases the PMSs cover at least two-thirds of the whole sheath, consistent with \cite{jones2002} and \cite{savani2011}. The PMS durations (average 6.0 hours) were also in agreement with the previous studies \citep{jones2000, nakagawa1993}. In addition, we found a significant amount of short PMSs (durations between 1-6 hours), as suggested by \citet{jones2000}.

Considering the two suggested PMS formation mechanisms (see Introduction), we focused on comparing differences between PMSs in the Near-Shock and Near-LE regions. For example, we found that the shock and PMS normals generally coincide for the Near-Shock group events, while the Near-LE group events had clearly larger deviations (Fig. \ref{fig:standoff}). We also found that in agreement with \cite{kataoka2005} the events in the Near-Shock group were associated with relatively strong (high $M_A$) shocks with low upstream $\beta$ (Figs. \ref{fig:histograms}a and \ref{fig:histograms}c). These shocks were also highly quasi-perpendicular, but a similar tendency was found for all investigated PMS-related groups.  According to \cite{kataoka2005}, downstream regions behind quasi-parallel and high upstream $\beta$ shocks become highly turbulent, which inhibits the formation of PMSs. A low upstream $\beta$ ahead of an IP shock would create suitable conditions for the alignment of the magnetic discontinuities as they pass the shock.

In addition, the cases where PMSs were observed near the CME leading edge had higher expansion speeds (Fig. \ref{fig:histograms}l) and higher difference between the CME leading edge and the ambient solar wind speeds (Fig. \ref{fig:histograms}k) than the events where PMSs occurred close to the shock. This is consistent with the amount of IMF draping ahead of the ejecta increasing with the increasing CME speed and expansion rate (e.g., \citealp{gosling1987}).  

Interestingly, we found PMSs most often within the Mid-Sheath region. There is no obvious (compressive) physical mechanism that would explain the formation of PMSs in the middle of the sheath. Hence, we argue that Mid-Sheath PMSs that have formed relatively early phase of the CME travel from Sun to Earth due to shock alignment and amplification. The Mid-Sheath-only events also stand-out in our statistics by having the largest plasma $\beta$ upstream of the shock (Fig. \ref{fig:histograms}c) and in the sheath (Fig. \ref{fig:histograms}g), the largest sheath thickness (Fig. \ref{fig:histograms}h) and being associated with the strongest shocks (Fig. \ref{fig:histograms}a) and the strongest CME expansion speeds (Fig. \ref{fig:histograms}k) and speed gradients (Fig. \ref{fig:histograms}l). It should be noted that as there were only nine such events, it is not possible to draw strong conclusions.

We found that plasma conditions in planar parts of the sheaths are generally in agreement with the previous studies (e.g., \citealp{nakagawa1989, nakagawa1993}): the PMS formation does not depend significantly on the local magnetic field magnitude nor on the local solar wind speed, while the plasma $\beta$ and density show clear variations. We also found that conditions associated with PMS formation depend on the location of the PMS, giving further support for distinct generation mechanisms in different parts of the sheath. Near the CME leading edge the plasma $\beta$ was clearly higher for planar cases (Fig. \ref{fig:histograms2}e). In turn, for the Near-Shock and Mid-Sheath regions the plasma $\beta$ was lower when PMSs were present. It should be noted that the plasma $\beta$ has a similar value for all PMS-cases, while larger differences appear in the non-planar parts of the sheath. This suggests that, near the CME leading edge, a low $\beta$ (and therefore a low level of compression in the plasma) inhibits the magnetic field from deflecting the plasma and draping its lines around the ejecta.  On the other hand, near the IP shock, a locally high $\beta$ (and therefore high compression) generates more turbulences and the discontinuities cannot be aligned into the two-dimensional plane. The relatively high densities in planar Near-LE regions (Fig. \ref{fig:histograms2}b) may be connected to some additional mechanism favoring the formation of PMSs, such as a piled-up compression region which forms during the early CME evolution (PUC, e.g., \citealp{das2011}), while lower densities in non-planar cases may indicate the presence of a plasma depletion layer (PDL, e.g., \citealp{liu2006}). 

The last part of our statistical analysis demonstrated that planar regions in sheaths are more likely to be geoeffective than non-planar regions. The planar regions, in particular in the middle of the sheath, exhibited significantly more strongly southward magnetic fields (Fig. \ref{fig:geo}). This is expected as both the shock compression and field line draping generate out-of-ecliptic magnetic fields.

In conclusion, the results presented here support two different PMS formation mechanisms, namely the amplification and alignment of solar wind discontinuities at the CME shock and the draping of the IMF around the ejecta. Moreover, the importance of compression mechanisms for PMS formation is highlighted near the CME leading edge. Such formation processes together with the relation between PMSs and strongly negative out-of-ecliptic fields may help in predicting sheath space weather effects at Earth. The correspondence between PMS and shock planes is promising in this regard.



\begin{acknowledgements}
We thank A. Szabo for the WIND MFI data, K.W. Ogilvie for the WIND SWE data, N. Ness for the ACE magnetic field data and  D.J. McComas for the ACE SWE data.\\
EP acknowledges the Magnus Ehrnrooth foundation for financial support. EK acknowledges Academy of Finland project 1267087 for financial support.\\
This paper uses data from the Heliospheric Shock Database, generated and maintained at the University of Helsinki.
\end{acknowledgements}

\end{document}